%% file: main.tex
\documentclass[sigconf]{acmart}
\usepackage[most]{tcolorbox}
\usepackage{tabularx}
\usepackage{booktabs}
\usepackage{multirow}
\usepackage{float}
\usepackage{subcaption}
\usepackage{caption}
\AtBeginDocument{%
  }

\setcopyright{acmlicensed}
\copyrightyear{2026}
\acmYear{2026}
\acmDOI{XXXXXXX.XXXXXXX}
\acmConference[Conference acronym 'XX]{Woodstock, NY}
\acmISBN{978-1-4503-XXXX-X/20--/--}

\begin{document}

\title{Eliciting Least-to-Most Reasoning for Phishing URL Detection}

\author{Holly Trikilis}
\email{htri0928@uni.sydney.edu.au}
\affiliation{%
  \institution{University of Sydney}
  \city{Sydney}
  \country{Australia}}
\author{Pasindu Marasinghe}
\email{pasindu.marasinghe@sydney.edu.au}
\affiliation{%
  \institution{University of Sydney}
  \city{Sydney}
  \country{Australia}}

\author{Fariza Rashid}
\email{fariza.rashid@sydney.edu.au}
\affiliation{%
  \institution{University of Sydney}
  \city{Sydney}
  \country{Australia}}

\author{Suranga Seneviratne}
\email{suranga.seneviratne@sydney.edu.au}
\affiliation{%
  \institution{University of Sydney}
  \city{Sydney}
  \country{Australia}}

\begin{abstract}
Phishing continues to be one of the most prevalent attack vectors, making accurate classification of phishing URLs essential. Recently, large language models (LLMs) have demonstrated promising results in phishing URL detection. However, their reasoning capabilities that enabled such performance remain underexplored. To this end, in this paper, we propose a Least-to-Most prompting framework for phishing URL detection. In particular, we introduce an ``answer sensitivity" mechanism that guides Least-to-Most's iterative approach to enhance reasoning and yield higher prediction accuracy. We evaluate our framework using three URL datasets and four state-of-the-art LLMs, comparing against a one-shot approach and a supervised model. We demonstrate that our framework outperforms the one-shot baseline while achieving performance comparable to that of the supervised model, despite requiring significantly less training data. Furthermore, our in-depth analysis highlights how the iterative reasoning enabled by Least-to-Most, and reinforced by our answer sensitivity mechanism, drives these performance gains. Overall, we show that this simple yet powerful prompting strategy consistently outperforms both one-shot and supervised approaches, despite requiring minimal training or few-shot guidance. Our experimental setup can be found in our Github repository \url{github.sydney.edu.au/htri0928/least-to-most-phishing-detection}.
\end{abstract}



\maketitle

\input{sections/introduction}
\input{sections/litreview}
\input{sections/methodology}
\input{sections/evaluation}
\input{sections/results}

\input{sections/conclusion}

\bibliographystyle{ACM-Reference-Format}
\bibliography{references}
\end{document}

%% file: sections/introduction.tex
\section{Introduction}

Phishing remains one of the most common cyber attacks, serving as one of the main entry points for many data breaches~\cite{accreport2024, ibmintelindex}. Existing detection methods include machine-learning and deep-learning methods, which analyse lexical, domain, and content-based features for classification~\cite{ozcan2023hybrid, alani2022phishnot}. More recently, URL-based approaches~\cite{urnet, catchphish} have enabled day-zero detection without requiring webpage content. These models perform well but generally lack interpretability, offering predictions without explanations for why a URL is classified as benign or phishing.

\begin{figure}[!htbp]
\centering
\includegraphics[width=0.7\columnwidth]{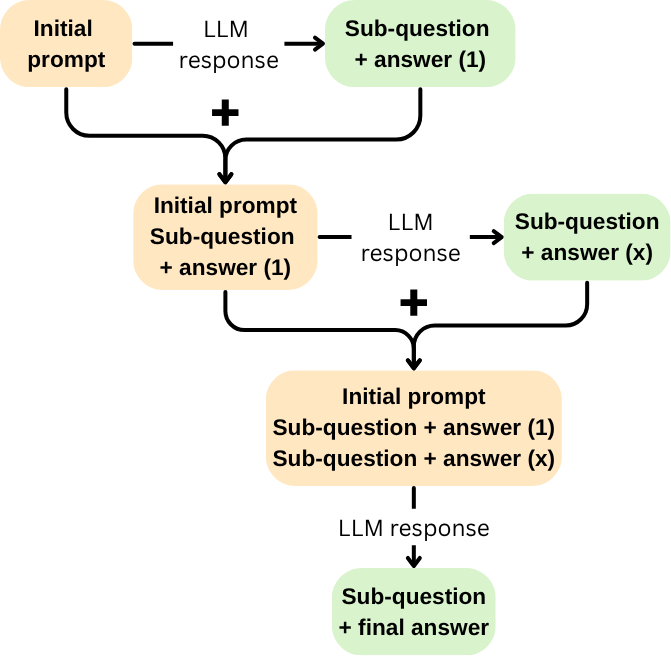}
\caption{General Least-to-Most prompting framework.}
\label{ltm figure}
\end{figure}

Large language models (LLMs) offer an interpretable alternative for classification tasks by leveraging their human-like reasoning capabilities~\cite{fewshot, cot}, and recent studies have applied these methods to phishing URL detection~\cite{prompt-v-tune, llm-oneshot}. While demonstrating high accuracy under few-shot settings, they do not analyse how their reasoning influences the final prediction. To address this gap, we propose LLM reasoning for phishing URL classification using Least-to-Most prompting~\cite{least-to-most}. As illustrated in Figure~\ref{ltm figure}, this strategy explicitly instructs the model to solve a given problem by generating smaller sub-problems to solve iteratively, gradually building toward a final prediction. This decomposition process provides deeper insight into the model’s internal reasoning compared with One-shot~\cite{llm-oneshot} or Chain-of-Thought~\cite{cot} prompting, which yield predictions in a single prompt or a single chain.

Our main contributions are the following:
(1) We propose a Least-to-Most-based prompt-engineering framework for phishing URL detection and introduce an answer sensitivity mechanism to guide the iterative reasoning process, improving the overall prediction accuracy.
(2) We evaluate our approach against two baselines—a one-shot prompt and the supervised URLTran model— and show that Least-to-Most delivers an average improvement of ~0.03 F1 over One-shot. Moreover, in its best-performing setting, our method comes within 0.03 F1 of URLTran’s performance.
(3) Through in-depth analysis, we show that Least-to-Most’s iterative process—reinforced by our answer sensitivity mechanism—effectively elicits stronger LLM reasoning, consistently improving performance over both one-shot and supervised approaches. Notably, it achieves these gains as a simple prompting technique that requires no few-shot examples or labelled training data.

%% file: sections/litreview.tex
\section{Related Work}

Phishing URL detection has been widely studied across blacklist-based, feature-engineered, and deep learning approaches. Early systems relied on curated blacklists such as Google Safe Browsing and PhishTank, but these are slow to update and ineffective against rapidly mutating phishing URLs \cite{blacklists}. To overcome these limitations, researchers developed content-based methods such as logo extraction and keyword analysis, along with lexical URL-only techniques such as PhishDef and bag-of-words URL representations \cite{tan,cantina,phishdef}. Deep learning models later demonstrated that URL strings alone provide strong discriminatory signals. CNN LSTM hybrids like MFPD~\cite{mfpd}, character and word-level CNN models such as URLNet~\cite{urnet}, and transformer-based models including BERT, ELECTRA~\cite{light-url} and URLTran~\cite{urltran} achieved state-of-the-art performance across multiple benchmark datasets. Despite their accuracy, these supervised models require large labelled datasets, offer limited interpretability, and lack generalization across datasets. 

Recently, LLMs provide an interpretable alternative for phishing detection by enabling classification through prompt engineering and in-context learning. Prompting strategies such as few-shot prompting, Chain of Thought (CoT) prompting and Least to Most decomposition have shown strong results across many reasoning-heavy tasks \cite{fewshot,cot,least-to-most}. Recent work has adapted these techniques to phishing, including email phishing classification \cite{email-phishing}, webpage-based detection \cite{website-lee,website-sasazawa,website-koide}, and URL-only phishing detection using few-shot or one-shot prompting \cite{opensource,llm-oneshot}. However, most LLM-based approaches use single-step prompts or treat the model’s reasoning as a black box. In contrast, our work applies Least to Most prompting~\cite{least-to-most} to iteratively decompose and refine URL analysis, showing that structured multi-step reasoning improves accuracy over earlier prompt-based methods while approaching the performance of supervised models such as URLTran \cite{urltran}.

%% file: sections/methodology.tex
\section{Least-to-Most Prompting for Phishing URL Detection} \label{chap:methodology}

As stated earlier, Least-to-Most prompting~\cite{least-to-most} addresses a question through sub-problem decomposition. While this is sufficient for standard reasoning tasks such as mathematical reasoning where it is easy to identify when an answer has been reached, we introduce a novel answer sensitivity mechanism to improve the prediction accuracy for phishing URL classification. That is, at each sub-question, we instruct the LLM  to provide a percentage estimate of the URL being phishing (0\% = benign, 100\% = phishing). This percentage estimate guides the iterative approach such that iterations only stop when the score crosses user-defined upper or lower sensitivity thresholds. For example, if the percentage estimate is equal to or above the upper sensitivity (therefore phishing), or equal to or lower than the lower sensitivity (therefore benign), the URL is considered classified. We programmatically limit the total number of iterations to 10 iterations, at which point the URL is conservatively considered phishing if the percentage estimate still has not crossed the upper or lower sensitivity thresholds. 

Figure~\ref{answer-sense-example} portrays an example of our approach. In the case of phishing URL detection, the LLM develops sub-questions and is guided by our answer sensitivity mechanism to assess the maliciousness of a URL, thereby giving insight into the specific reasoning steps yielding a final prediction. Compared to typical CoT prompting, which completes its breakdown in a single prompt, Least-to-Most is conducted over multiple outputs, using previous outputs to consider its next answer. 

\begin{tcolorbox}[colback=gray!10!white, colframe=black, title=Least-to-Most prompting with answer sensitivity example, %
    enhanced, 
    breakable,
    skin first=enhanced,
    skin middle=enhanced,
    skin last=enhanced,
    ]{}
\small

{\bf Prompt:} \\ \vspace{-4mm}

\texttt{\url{http://www.santonsofbendigo.com.au/install/Docs/document.php}}\\ \vspace{-2mm}

INSTRUCTIONS: \\
Identify if given URL is phishing or benign. Do not fully solve the problem. Begin by decomposing the question into one sub-question you need to solve in order to find the final answer. Solve this only question and do not continue. You MUST provide a prediction. Answer strictly in this form:\\ \vspace{-2mm}

\textit{Sub-question}: your sub-question here\\
\textit{Answer}: answer the sub-question in 80-100 words\\
\textit{Likeliness of phishing}: whole number percentage of how likely the URL is a phishing link (no other text is to be provided)\\ \vspace{-2mm}

Ensure a response is written every time \\ \vspace{-2mm}

{\bf Response:} \\ \vspace{-3mm}
        
\textit{Sub-question}: What is the primary domain of the URL, and does its structure or name immediately suggest brand impersonation or a common phishing tactic (e.g., typosquatting, suspicious TLD)? \\ \vspace{-3mm}

\textit{Answer}: The primary domain identified in the URL is `santonsofbendigo.com.au`. This domain structure, concluding with the `.com.au` country-code top-level domain, is standard for commercial entities...\\\vspace{-3mm}

\textit{Likeliness of phishing}: 35\\ \vspace{-2mm}

\textit{Sub-question}: What do the directory names (`/install/Docs/`) and the filename (`document.php`) within the URL's path suggest, particularly in the context of potential website compromise or hosting of malicious content?

\textit{Answer}: The directory structure `/install/Docs/` combined with the generic filename `document.php` raises immediate suspicion...

\textit{Likeliness of phishing}: 90

\end{tcolorbox}

\begingroup
\setlength{\abovecaptionskip}{2pt}   
\setlength{\belowcaptionskip}{0pt}   
\captionof{figure}[Least-to-Most prompting with answer sensitivity example.]%
{Prompt asking the model to break down into sub-questions of how to answer whether the given URL is phishing or benign based on the described answer sensitivity method.}
\label{answer-sense-example}
\endgroup
\vspace{-0.5em} 

%% file: sections/evaluation.tex
\section{Evaluation}\label{evaluation} 

To evaluate our framework, we ran experiments on four state-of-the-art models: Gemma 3:12b \cite{gemma}, Llama 3.1:8b \cite{llama}, GPT-4.1 \cite{OpenAI_2025}, and Gemini 2.5-Flash \cite{gemini}. Each model was evaluated using three publicly available URL datasets: HP~\cite{hp}, EBBU~\cite{ebbu}, and ISCX~\cite{iscx}. We compared our framework with two state-of-the-art models: \textbf{1) URLTran}~\cite{urltran} - a BERT-based model finetuned using labelled URL datasets, and \textbf{2) One-shot classifier}~\cite{llm-oneshot} - a one-shot prompting framework for phishing URL classification. We fine-tuned URLTran using the dataset splits from ~\cite{domain-adapt}. For testing all models, we extracted a random, balanced subset of 1,000 URLs from each of the three datasets mentioned above. Each experimental scenario was repeated five times, and the average F1 score is used for comparison. 

%% file: sections/results.tex
\section{Results and Analysis}

\begin{table}[H]
\centering
\scriptsize
\caption{F1 score comparison of Least-to-Most vs baselines.}
\setlength{\tabcolsep}{4pt}
\begin{tabular}{l l c c c}
\toprule
\textbf{LLM} & \textbf{Dataset} & \textbf{Least-to-Most} & \textbf{One-shot} & \textbf{URLTran} \\
\midrule

\multirow{3}{*}{\textbf{Gemma}} 
& HP   & 0.8872 & 0.8534 & 0.99 \\
& EBBU & 0.8962 & 0.8710 & 0.99 \\
& ISCX & 0.8283 & 0.8180 & 0.99 \\
\midrule

\multirow{3}{*}{\textbf{Llama}}
& HP   & 0.8763 & 0.7965 & 0.99 \\
& EBBU & 0.8580 & 0.8173 & 0.99 \\
& ISCX & 0.8033 & 0.7451 & 0.99 \\
\midrule

\multirow{3}{*}{\textbf{GPT}}
& HP   & 0.9502 & 0.9488 & 0.99 \\
& EBBU & 0.9490 & 0.9372 & 0.99 \\
& ISCX & 0.9133 & 0.9217 & 0.99 \\
\midrule

\multirow{3}{*}{\textbf{Gemini}}
& HP   & 0.9658 & 0.9612 & 0.99 \\
& EBBU & 0.9564 & 0.8910 & 0.99 \\
& ISCX & 0.9643 & 0.9110 & 0.99 \\
\bottomrule
\end{tabular}
\label{Tab:F1-compare}
\end{table}

In Table~\ref{Tab:F1-compare}, we report the mean F1 scores over five runs for each model–dataset pair. Overall, both Least-to-Most and One-shot prompting achieve performance comparable to URLTran, without the requirement of large labelled datasets. However, Least-to-Most generally outperforms One-shot, yielding an average F1 score of 0.9040 across all datasets and models, compared to 0.8726 for One-shot. This is enabled by Least-to-Most's iterative reasoning process for URL classification, thereby achieving higher prediction accuracy. Among the LLMs, Gemini performs best under Least-to-Most prompting with an average F1 score of 0.9621 across all datasets. This is only 0.0279 less than URLTran's supervised performance of a 0.99 F1 score. Next, we conduct an in-depth analysis of these results, focusing on the effectiveness of iterations and answer sensitivity, and the consistency of Least-to-Most's performance improvement over One-shot.

\subsection{Effect of iterations and answer sensitivity}
\begin{figure}[t]
    \centering
    
    \includegraphics[width=\linewidth]{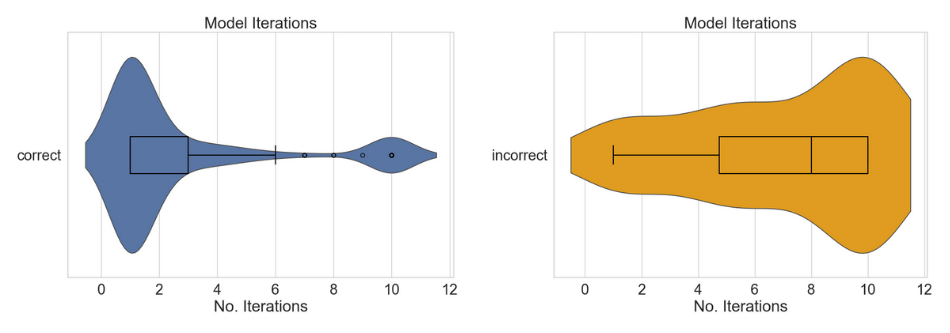}
    \caption[gpt-ebbu-boxplot]{Distribution of number of iterations (GPT-EBBU)}
    \label{gpt-ebbu-boxplot}
\end{figure}

We investigated the effectiveness of iterations by analysing the distribution of iterations for correct and incorrect predictions. Figure~\ref{gpt-ebbu-boxplot} exemplifies this for GPT on the EBBU dataset, and the patterns observed were similar across all scenarios. Overall, very few iterations were required for correct predictions, whereas the distribution for incorrect predictions was more varied. Interestingly, however, a significant number of correct predictions were enabled by an outlier number of iterations in most experiments. Isolating these specific cases, we investigated their trajectory of answer sensitivity values over the range of iterations.

\begin{figure}[H]
\centering
    
    \includegraphics[width=0.5\textwidth]{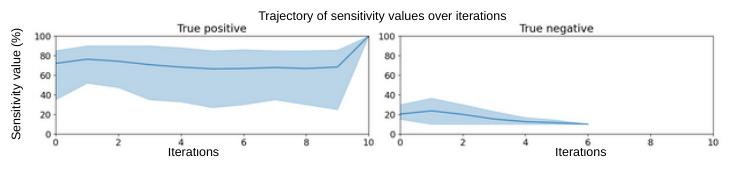}
\caption{Trajectory of sensitivity values (GPT-EBBU).}\label{agg-url}
\label{fig:likelyhood}
\end{figure}

 Figure~\ref{agg-url} portrays this for true positive and true negative cases of a random sample of 200 URLs (100 benign and 100 phishing URLs) taken from GPT results on the EBBU dataset, excluding single iteration predictions. The blue line represents the mean sensitivity value at a given iteration, and the shaded area represents the 10th to the 90th percentile. We can see that for many true positive cases, the sensitivity value begins below 50\%, and ends with a higher sensitivity giving the correct prediction by the final iteration (vice versa for true negatives). An example includes ``http://gargacharyasamaj.com.np/admin/oditole/mpot'', a phishing URL with an initial sensitivity of 25\% (a prediction closer to benign) which reached a sensitivity of 90\% (a correct phishing prediction) after six iterations. Without the iterations, this URL would have been incorrectly predicted.

Furthermore, the outlier number of iterations enabled our framework to improve the prediction accuracy for a significant number of URLs in comparison to One-shot~\cite{llm-oneshot}. We observed, for example, that for the EBBU dataset, out of the 24 URLs correctly predicted through outlier iterations by GPT, 19 had been incorrectly labelled by the single prompting method of One-shot. This pattern was repeated for all the other models. In most cases, the number of URLs that Least-to-Most corrects from One-shot through outlier iterations is at least half or more of the total URLs collected.

\subsection{Performance consistency}
Next, we investigated how consistently Least-to-Most outperformed One-shot in the regular, non-outlier iteration case. As shown in Table~\ref{tab:venn-correct}, for each experiment we enumerated the common URLs that both methods predicted correctly, as well as the number of unique URLs each method predicted correctly. In almost all cases, the number of unique correct predictions by Least-to-Most exceeded that of One-shot. Alternatively, in almost all cases, Least-to-Most made fewer incorrect predictions than the One-shot method. In extreme cases, such as Llama, the difference in both amounts is over 100. This fact stands out as Llama was the poorest performing model for Least-to-Most, showing that even in its worst case that it improves One-shot results. This significant gap shows how consistently the Least-to-Most method improves upon the One-shot method. Further details about all our analysis findings can be found in our Github repository.
\vspace{-2mm}

\begin{table}[h]
\centering
\scriptsize
\caption{Least-to-Most vs One-shot: Correct predictions.}
\label{tab:venn-correct}

\begin{tabular}{l l c c c}
\toprule
\textbf{LLM} & \textbf{Dataset} &
\textbf{Least-to-Most correct} & \textbf{Both correct} & \textbf{One-shot correct} \\
\midrule

\multirow{3}{*}{Gemma}
 & HP   & 63 & 791 & 41 \\
 & EBBU & 60 & 790 & 20 \\
 & ISCX & 57 & 739 & 47 \\
\midrule

\multirow{3}{*}{Llama}
 & HP   & 158 & 700 & 45 \\
 & EBBU & 155 & 723 & 60 \\
 & ISCX & 162 & 616 & 56 \\
\midrule

\multirow{3}{*}{GPT}
 & HP   & 19 & 928 & 19 \\
 & EBBU & 25 & 919 & 18 \\
 & ISCX & 35 & 911 & 8 \\
\midrule

\multirow{3}{*}{Gemini}
 & HP   & 16 & 945 & 16 \\
 & EBBU & 72 & 882 & 14 \\
 & ISCX & 65 & 900 & 11 \\
\bottomrule
\end{tabular}
\end{table}

%% file: sections/conclusion.tex
\section{Conclusion} \label{chap:conclusion}

In this work, we propose a Least-to-Most prompting framework for phishing URL detection. We show that its iterative reasoning process, combined with our answer sensitivity mechanism, enables enhanced LLM reasoning and higher prediction accuracy than baseline methods. On average, our approach outperforms the One-shot baseline by about 0.03 F1, and our best-performing LLM comes within 0.03 F1 of the state-of-the-art supervised model, URLTran. Our analysis further reveals that iterations guided by answer sensitivity correct a substantial number of predictions that are initially incorrect. Overall, our method consistently surpasses the One-shot baseline while requiring no labelled data or few-shot examples.